\documentstyle [aps,pra]{revtex}

\newcommand {\beq}{\begin{equation}}
\newcommand {\eeq}{\end{equation}}
\newcommand {\la} {\langle}
\newcommand {\ra} {\rangle}

\title{Quantum Zeno and anti-Zeno effects in the Friedrichs model}
\author{I. Antoniou$^1$, E. Karpov$^1$, G. Pronko$^{1,2}$, and 
E.~Yarevsky$^{1,3}$}
\address{$^1$ International Solvay Institutes for Physics and Chemistry,
C.P. 231, Campus Plaine ULB, Bd. du Triomphe, Brussels 1050, Belgium}
\address{$^2$ Institute for High Energy Physics, Protvino, Moscow region 
142284, Russia}
\address{$^3$ Laboratory of Complex Systems Theory, Institute for Physics,
St.Petersburg State University, Uljanovskaya~1, St.Petersburg 198904, Russia}

\begin{document}

\maketitle
\begin {abstract}
We analyze the short-time behavior of the survival probability in the frame
of the Friedrichs model for different formfactors. We have shown that this
probability is not necessary analytic at the time origin. The time when 
the quantum Zeno effect could be observed is found to be much smaller than 
usually estimated. We have also studied the anti-Zeno era and have estimated 
its duration.
\end {abstract}


\section {Introduction}

Since the very beginning of the quantum mechanics, the measurement 
process has been a most fundamental issue. The main characteristic 
feature of the quantum measurement is that the measurement changes the
dynamical evolution. This is the main difference
of the quantum measurement compared to its classical analogue. On this 
framework, Misra and Sudarshan pointed out~\cite {Misra} that repeated 
measurements can prevent an unstable system from decaying. Indeed,
as the survival probability is in most cases proportional to the square 
of the time for short times (see, however, the discussion below), the 
measurement effectively projects the evolved state back to the initial state
with such a high probability that the sequence of the measurements 
``freezes'' the initial state. Led by analogy with the Zeno paradox, this 
effect has been called the quantum Zeno effect (QZE).

Cook~\cite {sugg-exp} suggested an experiment on the QZE which was 
realized by Itano et al.~\cite {experiment}. In this 
experiment, the Rabi oscillations have been used in order to demonstrate
that the repeated observations slow down the transition process. However,
the detailed analysis~\cite {PPT,disc9,disc91} has shown that the results 
of this experiment could equally well be understood using a density matrix 
approach for the whole system. Recently, an experiment similar 
to~\cite {experiment} has been performed by Balzer et al.~\cite {Toschek} 
on a single trapped ion. This experiment has removed some drawbacks usually
associated with the experiment of Itano et al.~\cite {experiment}, for 
example, dephasing
system's wave function caused by a large ensemble and non-recording
of the results of the intermediate measurements pulses. We refer to recent
reviews~\cite {Whitaker,quantph} for detailed discussions of these and 
related questions.

Both experiments~\cite {experiment,Toschek} demonstrate the perturbed 
evolution of a coherent dynamics, as opposed to spontaneous decay. 
So the demonstration of the QZE for an unstable 
system with exponential decay, as originally proposed in~\cite {Misra}, 
is still an open question. The main problem in such an experimental 
observation of the Zeno effect is the very short time when the quadratic
behavior of the 
transition amplitude is valid~\cite {P1,P2}. On the other hand, the 
Zeno-type experiment could also reveal deviations from the exponential 
decay law and the magnitude of these deviations.

The QZE has been discussed for many physical systems including 
atomic physics~\cite {P1,P2,atom1,atom2}, radioactive decay~\cite {radio},
mesoscopic physics~\cite {mesoscopic1,mesoscopic2,mesoscopic3,mesoscopic4},
and has been even proposed as a way to control decoherence for effective
quantum computations~\cite {decoherence}. Recently, however, a quantum 
anti-Zeno effect has been found~\cite {antiZeno1,antiZeno2}.
Under some conditions the repeated observations could speed up the decay 
of the quantum system. The anti-Zeno effect has been further analyzed 
in~\cite {mesoscopic4,PRA61-022105,PRA61-052107,Nature,Pasc-prepr}. 

We carefully analyze here the short-time behavior of the 
survival probability in the frame of the Friedrichs model~\cite {Fried}.
We have shown that this probability is not necessary analytic at zero time.
Furthermore, the probability may not even be quadratic for the short times 
while the QZE still exists in such a case~\cite {antiZeno1,Misra2}. 
We have shown (see also Kofman and Kurizki~\cite {Nature}) that the time 
period within which the QZE could be observed is much smaller than 
previously believed. Hence we conclude that the experimental 
observation/realization of the QZE is quite challenging.

We have also analyzed the anti-Zeno era. While it seems that most
decaying systems exhibit anti-Zeno behavior, our examples contradict the 
estimations of Lewenstein and Rzazewski~\cite {PRA61-022105}. 
We have studied the duration of the anti-Zeno era and have estimated 
this duration when possible.

\section {Model and exact solution}

The Hamiltonian of the second quantised formulation of the Friedrichs 
model~\cite{Fried} is
\beq
  H =  H_0 + \lambda V, 
\label{H}
\eeq
where the unperturbed Hamiltonian is defined as 
$$
  H_0  =  \omega_1a^\dag a
        + \int^{\infty}_0\!\!d \omega\,\omega\,b^\dag_\omega b_\omega ,
$$
and the interaction is 
\beq
\label {Ham2}        
  V = \int_0^{\infty}\!\!d\omega f(\omega)
          \left(ab^\dag_\omega + a^\dag b_\omega \right).
\eeq
Here $a^\dag$, $a$ are creation and annihilation boson operators of the atom 
excitation, $b^\dag_\omega$, $b_\omega$ are creation and annihilation boson 
operators of the photon with frequency $\omega$, $f(\omega)$ is the 
formfactor, $\lambda$ is the coupling parameter, and the vacuum energy is 
chosen to be zero. The creation and annihilation operators satisfy the 
following commutation relations:
\equation \label{cra}
  [a, a^\dag]
    = 1 , \qquad   
  [b_\omega, b^\dag_{\omega'}]
    = \delta(\omega - \omega').
\endequation
All other commutators vanish.

The Hamiltonian $H_0$ has continuous spectrum $[0,\infty)$ of uniform
multiplicity, and the discrete spectrum 
$n\omega_1$ (with integer $n$) is embedded in the continuum. 
The space of the wave functions is the direct sum of the Hilbert 
space of the oscillator and the Fock space of the field.

For $\omega_1 > 0$ the oscillator excitations are unstable due to the 
resonance between the oscillator energy levels and the energy of a photon.
Therefore, the total evolution leads to the decay of a wave packet 
corresponding to the bare atom $|1\rangle$. Decay is described by the 
survival probability $p(t)$ to find, after time $t$, the bare atom evolving 
according to the evolution $\exp(-iHt)$ in its excited state~\cite {P2}:
\beq
\label {survprobdef}
  p(t) \equiv |\langle 1|e^{-iHt}|1\rangle|^2.
\eeq
The survival probability can be easily calculated in the second quantized 
representation:
$$
p(t) = |\langle 0 | a(0) e^{-iHt} a^\dag(0) |0 \rangle|^2 =
|\langle 0 | e^{-iHt} e^{iHt} a(0) e^{-iHt} a^\dag(0) |0 \rangle|^2 =
|\langle 0 | a(t) a^\dag(0) |0 \rangle|^2,
$$
where $a^\dag(0)=a^\dag$. The time evolution of $a(t)$ in the Heisenberg
representation is presented in Appendix~A. Using (\ref{abt}) we obtain
$$
p(t) = |A(t)|^2,
$$
where the survival amplitude $A(t)$ is given by (\ref{A}).

Due to the dimension argument, we can write the formfactor 
$f(\omega)$ in the form
$$
f^2(\omega) = \Lambda \varphi\left({\omega \over \Lambda}\right),
$$
where $\varphi(x)$ is a dimensionless function. Here $\Lambda$ 
is a parameter with the dimension of $\omega$.
The survival amplitude $A(t)$ in the dimensionless representation is
\beq\label{Adl}
  A(t) = \frac{1}{2\pi i} \int\limits_{-\infty}^\infty  dy
    \frac{e^{iy\Lambda t}}{\eta^-_\Lambda(y)},
\eeq
where
\beq
  \eta^-_\Lambda(z)=\omega_{\Lambda} -z - \lambda^2
  \int\limits_0^\infty  dx \frac{\varphi(x)}{x - z + i0},
\eeq
and $\omega_{\Lambda} = \omega_{1}/\Lambda$.

\section {Short-time behavior and the Zeno region}

The short time evolution of the model~(\ref{H}) depends
essentially on the formfactor. In order to illustrate different 
types of the evolution, we shall consider two formfactors, namely:
\beq
f_1^2(\omega) = \Lambda {\sqrt{\omega \over \Lambda} \over
1 + {\omega \over \Lambda} }, \quad
\varphi_1(x)={\sqrt{x} \over 1+x},
\eeq
and
\beq
f_2^2(\omega) = \Lambda {{\omega \over \Lambda} \over
\left(1 + \left({\omega \over \Lambda}\right)^2 \right)^2 } ,
\quad
\varphi_2(x)={x \over (1+x^2)^2}.
\eeq
The formfactor $f_1$ permits exact 
calculations~\cite {photodetachement1,kofman-opt}.
It turns out that the short-time behavior is not 
quadratic~\cite {antiZeno1,Misra2} as anticipated by~\cite {PRA61-022105}.
We shall also use for 
comparison the results presented in~\cite{P1,P2} for the formfactor 
$\varphi_3(x)={x \over (1+x^2)^4}$~\cite{reffromPasc}. To get a first 
impression about the time scales, we associate with each formfactor a 
physical system: the photodetachement process for 
$\varphi_1(x)$~\cite{PRA61-022105,photodetachement1,photodetachement2}, 
the quantum dot for $\varphi_2(x)$~\cite{PRB}, and the hydrogen atom for
$\varphi_3(x)$~\cite{P1}. The corresponding numerical values of the 
parameters $\Lambda$, $\omega_1$, and $\lambda^2$ are listed in Table~1.
We would like to emphasize that these values (as well as the model itself)
are approximate estimations of the corresponding effects.

Let us discuss the short-time behavior of the survival probability $p(t)$. 
We shall assume here the existence of all necessary matrix elements, and 
denote $\la \cdot \ra  = \la 1|\cdot|1 \ra$.
\begin{eqnarray}
 p(t) = \la e^{-iHt} \ra = \la 1-iHt-{1\over 2}H^2t^2 + {i\over 6}H^3t^3+
{1\over 24}H^4t^4+O(t^5) \ra = \nonumber \\
 \left( 1-{t^2\over 2} \la H^2 \ra+{t^4\over 24} \la H^4 \ra\right)^2 +
\left( t \la H \ra -{t^3\over 6} \la H^3 \ra\right)^2+O(t^6) = \nonumber \\
 1-t^2 \left(\la H^2 \ra - \la H \ra^2\right) + 
t^4 \left({1\over 4} \la H^2 \ra^2 +{1\over 12} \la H^4 \ra -
{1\over 3} \la H \ra \la H^3 \ra \right) +O(t^6) =  \nonumber \\
1 - {t^2\over t_a^2} + {t^4\over t_b^4} + O(t^6).
\label{p-expansion}
\end{eqnarray}
In order to calculate the parameters $t_a$ and $t_b$, we need to 
calculate the  averages of the powers of $H=H_0+\lambda V$:
\begin{eqnarray}
\label{labelM}
\la H \ra     & = & \omega_1, \nonumber \\
\la H^2 \ra & = & \omega_1^2 +\lambda^2 \la V^2 \ra, \nonumber \\
\la H^3 \ra & = & \omega_1^3 +2\lambda^2\omega_1 \la V^2 \ra 
+\lambda^2 \la VH_0V \ra, \\
\la H^4 \ra & = & \omega_1^4 +\lambda^2
\left(3\omega_1^2 \la V^2 \ra+2\omega_1 \la VH_0V \ra 
+ \la VH_0^2V \ra \right)+ \lambda^4 \la V^4 \ra. \nonumber
\end{eqnarray}
These expressions are valid in our model because of the special structure of 
the potential $V$~(\ref{Ham2}). Now we can find:
\begin{eqnarray}
\label {tz-gen}
{1\over t_a^2} & = & \lambda^2 \la V^2 \ra =
\lambda^2 \Lambda^2 I_0, \nonumber \\
{1\over t_b^4} & = & \lambda^2
\left({\omega_1^2 \over 12} \Lambda^2 I_0 -
{\omega_1 \over 6} \Lambda^3 I_1 + {\Lambda^4\over 12} I_2\right)
+\lambda^4 \Lambda^4 \left({ I_0^2 \over 4} 
+ { \int_0^\infty \varphi^2(x) dx \over 12}\right),
\end{eqnarray}
where
$$
I_k = \int_0^\infty x^k \varphi(x) dx.
$$
In the weak coupling models the following inequalities 
are satisfied (see Table~1):
\beq
\label {labelA}
\lambda^2 \ll 1 \quad \mbox{and}\quad \Lambda \gg \omega_1.
\eeq
In this approximation we can simplify the expression for $t_b$:
\beq
{1\over t_b^4}  \approx 
{\lambda^2 \Lambda^4 \over 12} I_2 .
\eeq

The parameter $t_a$ has been called Zeno time~\cite {P1,P2} because it has 
been conjectured to be related to the Zeno region, i.e. the region where 
the decay is slower than the exponential one and the Zeno effect can
manifest. On the other hand, a more precise estimation reveals that
the Zeno region is in fact orders of magnitude shorter than $t_a$. We 
illustrate this in Fig.~1 where the survival probabilities for the 
formfactors $\varphi_1(x)$ and $\varphi_2(x)$ are plotted. The 
corresponding analytical expressions and the numerical values for 
different time scales are presented 
in Table~1. We see that $p(t)$ is not convex already at times 
much shorter than the time $t_a$. 

In view of this fact, we propose another definition for the Zeno 
time. As one refers in discussions about the Zeno effect on the expansion 
of survival probability for small times, and specifically on the second term,
we shall define the Zeno time $t_Z$ as corresponding to the region where the 
second term dominates. Hence the introduced time $t_Z$ is a natural 
boundary where the second and third terms have the same amplitude:
\beq
\label {tr-gen}
{t_Z^2 \over t_a^2} = {t_Z^4 \over t_b^4}, \quad\mbox{so}\quad
t_Z=t_b^2/t_a.
\eeq
In the weak coupling models, 
$$
t_Z = {1 \over \Lambda} \sqrt{12 I_0 \over I_2},
$$ 
that agrees with the estimation in~\cite {Nature}.
This time is much shorter than $t_a \sim {1\over \lambda\Lambda}$ and agrees 
much better with the numerical estimations. For example, for the interaction 
$\varphi_3(x)$ we find 
$t_Z={2\sqrt{6} \over \Lambda} \approx 5.8\cdot 10^{-19}$ s while
$t_a={\sqrt{6} \over \lambda\Lambda} \approx 3.6\cdot 10^{-16}$ s~\cite{P1}. 

Our conclusions are in fact valid for a rather wide 
class of interactions. Namely, they are valid if the matrix 
elements~(\ref{labelM}) exist and conditions~(\ref{labelA}) are satisfied. 
For example, any bounded locally integrable interaction $\varphi(x)$ 
decreasing as $\varphi(x) \sim {C\over x^{1.5+\epsilon}}$, $\epsilon >0$ at 
$x\to\infty$, gives finite matrix elements~(\ref{labelM}). Furthermore, we 
show below that the relation $t_R \ll t_Z$ may be valid even when the matrix
elements~(\ref{labelM}) do not exist. 

For the formfactor $\varphi_1$, already the matrix element $\la V^2 \ra$ 
does not exist, and the short time expansion is written, Appendix~B, as
\[
p(t) = 1 - \left({t \over t_a}\right)^{1.5}
+\left({t \over t_b}\right)^2 +O(t^{5/2}),
\]
where $t_a = ( 3/(4\sqrt{2\pi}))^{2/3}/( \lambda^{4/3} \Lambda)$, and 
$t_b = 1/(\sqrt{\pi} \lambda \Lambda)$.
In fact, from the representation~(\ref{labelM}) 
one can easily deduce that for any formfactor decreasing according to the 
power law when $x \to \infty$, $p(t)$ is not analytic at $t=0$. 
Specifically, if the formfactor decreases as $\varphi(x) \sim x^\alpha$ 
when $x \to \infty$, only the Taylor coefficients up to $t^n$ with 
$n < 1+|\alpha|$ can be defined. 

Following the previous discussion, for the formfactor $\varphi_1(x)$ the 
Zeno time $t_Z$ can be estimated by the condition
$$ 
\left({t_Z \over t_a}\right)^{1.5} =\left({t_Z \over t_b}\right)^2,
\quad\mbox{so}\quad t_Z={32\over 9\pi\Lambda}.
$$
For this case, one can see that the time $t_a$ has scaling properties 
which differ from~(\ref{tz-gen}) while the Zeno time $t_Z$ has a value 
similar to~(\ref{tr-gen}). 

For $\varphi_2$, the matrix element $\la V^2 \ra$ exists so the usual 
time $t_a$ can be introduced: $t_a={\sqrt{2} \over \lambda\Lambda}$. 
However, $\la VH_0^2V \ra$ does not exist and the asymptotic behavior 
of $p(t)$ is (see Appendix~C)
\[
p(t)=1 - \left({t \over t_a}\right)^2-{\lambda^2 \over 12}\log{(2\omega_1 t)}
\Lambda^4 t^4+O(t^4).
\]
Repeating the arguments concerning the Zeno region, we find $t_Z = 
{\sqrt{6}\over \Lambda\sqrt{|\log{({2\sqrt{6}\omega_1 \over \Lambda})}|}}$.
In this case one can see again that the Zeno time $t_Z$ has a value 
similar to~(\ref{tr-gen}), and the inequality $t_Z \ll t_a$ is satisfied.

\section {Zeno and anti-Zeno effects}

The probability that the state $|1\ra$ after $N$ equally spaced measurements
during the time interval $[0,T]$ has not decayed, is given by~\cite {Misra}
\beq
\label {pnt}
p_N(T) = \la 1 | \left(|1\ra \la 1 | e^{-iHT/N}\right)^N |1 \ra = p^N(T/N) 
\la 1|1 \ra = p^N(T/N).
\eeq
Expression~(\ref{pnt}) is only correct for the ideal von Neumann
measurements~\cite {Misra}. We are interested in the behavior of $p_N(T)$ 
as $N \rightarrow \infty$ or, equally, when the time interval between 
the measurements $\tau=T/N$ goes to zero:
\beq
\label {Zenoas}
\lim\limits_{\tau \rightarrow 0} p_N(T)  = \lim\limits_{\tau \rightarrow 0} 
p(\tau)^{T/\tau}  = \left( \lim\limits_{\tau \rightarrow 0} 
\left(1-{{1-p(\tau) \over \tau}\over {1\over\tau} }\right)^{1\over \tau}  
\right)^T = \left\{\begin{array} {ll} 
0, & {\rm when} \quad p'(0) = -\infty, \\
e^{-cT}, & {\rm when} \quad p'(0) = -c, \\
1, & {\rm when} \quad  p'(0) = 0.
\end{array}  \right.
\eeq

Hence for the case $p(t)=1-ct^\alpha$ one has the Zeno effect for all 
$\alpha > 1$ \cite {antiZeno1,Misra2}. We should notice that in case of 
the linear asymptotics 
of $p(t)$ at short times (in particular, for the purely exponential decay) 
there is no Zeno effect, and the probability to find the system in the 
initial state $|1\ra$ decreases exponentially with the time of observation.
The results~(\ref{Zenoas}) are found in case of continuously ongoing
measurements during the entire time interval $[0,T]$. Obviously, this is 
an idealization. In practice we have a manifestation of the Zeno effect,
if the probability~(\ref{pnt}) increases as the time interval $\tau$
between measurements decreases. Formula~(\ref{Zenoas}) may be accepted as
an approximation for a very short time interval $\tau \ll t_Z$. For 
longer times we cannot use the Taylor expansion, therefore 
Eq.~(\ref{Zenoas}) is not valid. It appears that in order to analyze 
longer time behavior, 
the long time asymptotics of the $p(t)$ are more convenient. These 
asymptotics can be summarized as follows (see Appendicies~B, C):
\beq
p(t) \approx  |A_1|^2 e^{-4\gamma \sqrt{\omega_1\Lambda}t}+
{\pi \lambda^4 \Lambda \over 4 \omega_1^4 t^3} h_1^2(t) -
{\sqrt{\pi} \lambda^2 \Lambda^{1/2} \over \omega_1^2 t^{3/2}}
|A_1| h_1(t)
e^{-2\gamma \sqrt{\omega_1\Lambda}t} \cos{(\omega_1t - \pi/4)}
\label {longas1}
\eeq
when $t \gg 24/\omega_1$ for the $\varphi_1(x)$, and 
\beq
p(t) \approx |A_2|^2 e^{-\gamma_1 \Lambda t} + 
{\lambda^4\over \omega_1^4 t^4 } h_2^2(t)
 - {2 \lambda^2 e^{-\gamma_1 \Lambda t/2}\over \omega_1^2 t^2 }
|A_2|h_2(t) \cos{(\omega_1 t)}
\label {longas2}
\eeq
when $t \gg 4/\omega_1$ for the $\varphi_2(x)$. Here the constants 
$A_1$, $A_2$ satisfy the inequality $|1-|A_k|^2| \ll 1$, $k=1,2$.
The functions $h_1$, $h_2$ have the following asymptotic properties: 
$$
\lim\limits_{t \to \infty} h_1(t) = 1; \quad
\lim\limits_{t \to 0} {h_1(t) \over t^{3/2}} = const; \quad
\lim\limits_{t \to \infty} h_2(t) = 1; \quad
\lim\limits_{t \to 0} {h_2(t) \over t^2} = const.
$$
In paper~\cite{P1}, an expression very similar 
to~(\ref{longas2}) was found for the formfactor $\varphi_3(x)$. 
Expressions~(\ref{longas1},\ref{longas2}) are analytically 
established only in the region $t \gg C/\omega_1$. However, the numerical 
investigation shows that for our choice of parameters we can use
(\ref{longas1},\ref{longas2}) for a qualitative description already in the 
region $t \sim 1/\omega_1$. Then one can see that the oscillation with the 
frequency $\omega_1$ starts always with the negative cosine wave. Therefore, 
the survival probability~(\ref{survprobdef}) turns out to be
less than purely exponential, and one can expect decreasing of the 
probability $p_N(T)$ as well. We illustrate this effect in Fig.~2 for both 
the photodetachement process and the quantum dot. The anti-Zeno 
region (AZ region), i.e. the region where the probability $p_N(T)$ is less 
than purely exponential, is clearly seen for both systems. 
For $\tau \to 0$, $p_N(T)$ approaches $1$ according to~(\ref{Zenoas}).

We should stress that the above described behavior shows that the initial 
quadratic behavior is not just a beginning of the first wave of oscillation
as stated in~\cite {P1}. This is true because the time $t_a$ is actually not
the time within which $p(t)$ has quadratic behavior. In fact, the
quadratic behavior is only valid for $t \ll t_Z$ and has nothing in common 
with the oscillations in Eqs.~(\ref{longas1},\ref{longas2}).

On the basis of Fig.~2 we would like to make some additional remarks. First
of all, for lager observation time $T$, the AZ region is wider and
the probability $p_N(T)$ in the AZ region is lower. This is natural: 
the bigger the time of observation is, the harder to restore the initial 
state of the system. Secondly, one can see that the value $t_Z$ describes 
very well a minimum of the probability in the AZ region. For shorter times, 
$p_N(T)$ increases, but it still may be much less compared 
to the $p_N(T)$ in the purely exponential region. Hence, the classical
Zeno effect~\cite{Misra} could be observed only when $t \ll t_Z$.
Finally, we also notice that one can sometimes observe the second 
wave of oscillation in~(\ref{longas1},\ref{longas2}) (see Fig.~2a).
However, its amplitude is much less than the amplitude of the first wave.

A general consideration of the AZ region is presented 
in ~\cite {PRA61-022105}. The authors conclude that the AZ region exists 
for all generic weakly coupled decaying systems. Under some assumptions, 
they have found that 
\beq
\label {A-cond}
|A_k|^2 < 1,
\eeq
and use this condition for the explanation of the existence of the AZ region.
However, some assumptions made in~\cite{PRA61-022105} for the derivation 
of~(\ref{A-cond}) are not always valid. 
For example, for the model $\varphi_1(x)$ (the formfactor used 
in~\cite{PRA61-022105}) one calculates $|A_1|^2 \approx 1+1.1\ 10^{-6} > 1$ 
for our choice of the parameters. For the model $\varphi_2(x)$ we have 
$|A_2|^2 \approx 1+\lambda^2(3+2\log{\omega_\Lambda}) < 1$, but this effect 
is of the second order in the coupling while in~\cite {PRA61-022105} the 
fourth order was found. Hence the above mentioned results can not be 
considered as a proof of the existence of the AZ region.

Indeed, our results show that there exist two different types 
of the AZ region. The first case takes place as the amplitude
of oscillations in~(\ref{longas1},\ref{longas2}) is less than $|1-|A|^2|$,
and $|A|^2 <1$. This corresponds to the arguments of~\cite {PRA61-022105}.
In this situation, the survival probability is always less than the 
``ideal'' one corresponding to the pure exponential decay (except for the 
very short times $t \ll t_Z$). The second case arises
when the amplitude of oscillations in~(\ref{longas1},\ref{longas2}) is bigger 
than $|1-|A|^2|$ (for any $|A^2|$), or when $|A^2|>1$. In this case the 
survival probability may be lower or higher than the ``ideal'' one, 
that may result in oscillations of the probability $p_N(T)$. This is exactly
the situation in Fig.~2a.

It would be very interesting to find an estimation for the duration of the 
AZ region. We have found that the minimum of the $p_N(T)$ is 
reached at $t_Z$, however the whole region is much wider. Unfortunately, 
we can present this estimation only for the second type of the AZ region. 
In order to illustrate this, we plot in Fig.~3 the value
\beq
\label {Neps}
N_\varepsilon(T) : p_{N_\varepsilon(T)}(T) = (1-\varepsilon) p_1(T).
\eeq
This value gives the maximum number of repeated observation such that
the probability $p_N(T)$ would not be less than $p(T)$ with
accuracy $\varepsilon$. The difference between two types of the
AZ region is very pronounced. For the first type 
($\varphi_2(x)$ interaction) $N_\varepsilon(T) \sim C \varepsilon $ and
is almost independent of the time $T$ of observation. It means that the
anti-Zeno region $t_{AZ}$ should be described as 
$t_{AZ} \sim c T /\varepsilon$. So the duration depends critically
on the time of observation and the accuracy, and cannot be attributed to
the properties of the system itself. 

For the second type 
($\varphi_1(x)$ interaction) $N_\varepsilon(T) \sim C T$ and is almost
independent of the accuracy $\varepsilon$. This means that $t_{AZ}$ is
independent of the time of the observation and the accuracy, so
it can be correctly introduced. In fact, in this case $t_{AZ}$ is defined
by the oscillations of the survival probability and can be estimated 
as $1/\omega_1$.

The estimation $t_{AZ} \ll 1/\omega_1$ was given by Kofman and 
Kurizki~\cite {Nature}. While this estimation obviously holds, it is
necessary to establish more precise boundaries for $t_{AZ}$. 
We have found the boundary $t_{AZ} \sim 1/\omega_1$ for the 
$\varphi_1(x)$ interaction. 
However, from the results presented in Fig.~3, one can see that for the 
interactions $\varphi_2(x)$ and $\varphi_3(x)$ the estimation $1/\omega_1$
can hardly be used, contrary to the results of~\cite {Nature}.

We would like to mention that the estimation $t_{DC}=1/\omega_1$ has 
been obtained by Petrosky and Barsegov~\cite {ZenoTomio} as an upper boundary 
of the decoherence time marking the onset of the exponential era.
As the Zeno effect cannot be realized for times lager than $t_{DC}$, 
Petrosky and Barsegov called $t_{DC}$ the Zeno time. In fact this is 
a rough estimation of the real Zeno time $t_Z$.

\section {Conclusions}

Let us summarize the short-time behavior of the survival probability. 
We introduce two regions: the very short Zeno region $t_Z$ with the scale 
$1/\Lambda$ and the much longer anti-Zeno region $t_{AZ}$. If one performs 
a Zeno-type experiment, and the time between measurements is much shorter 
than $t_Z$, then the Zeno effect -- increasing of the survival probability --
can be observed. In the time range between $t_Z$ and $t_{AZ}$, the anti Zeno
effect exists, i.e. decay is accelerated by repeated measurements. That is 
why the Zeno time cannot be longer than $t_Z$. The previous estimations 
of the Zeno time $t_a$~\cite {P1,P2} and $t_{DC}$~\cite {ZenoTomio} are
much longer than our estimation $t_Z$ for physically relevant 
systems~(\ref{labelA}).

While the acceleration of decay is clearly seen in all cases, it is not 
always possible to introduce the value $t_{AZ}$. The reason is the possible
dependence of $t_{AZ}$ on the moment of the observation and on the accuracy 
of the observation. When this dependence is absent, one finds 
$t_{AZ} \sim 1/\omega_1$. Hence the anti-Zeno region is, for typical values 
of parameters, a few orders of magnitude longer than the Zeno region. 
It would be very important from the experimental point of view, to find 
an estimation for the anti-Zeno region in terms of the initial
parameters without any reference to the constant $A_k$.

It is possible in principle that the oscillations
in~(\ref{longas1},\ref{longas2}) may give a few successive Zeno and 
anti-Zeno regions. 
However, as the amplitude of the oscillations decreases exponentially 
with time, these regions are hardly visible. After the anti-Zeno region,
the system decays exponentially up to the time $t_{ep}$ when the long-tail 
asymptotics substitutes the exponential decay. 

In accordance with this picture, the experimental observation of the Zeno 
effect is very difficult. Indeed, the Zeno region appears to be considerably
shorter than previously belived. The 
acceleration of the decay should be observed before the deceleration will
be possible. In this connection, the proposals for using the Zeno effect
for increasing of the decoherence time~\cite {decoherence} should be 
critically analyzed. We conclude that the Zeno effect may not be very 
appropriate for decoherence control desired for quantum computations.

There seems to be no place for the usual estimations of the 
Zeno time by $t_a$. There are no physical effects which can be associated 
with this time scale.  In our opinion, the widespread expectation that the 
time $t_a$ describes the Zeno region, is based on a naive perturbation 
theory. One could assume that $p(t)=1-\sum_{k=2}^\infty c_k (\lambda t)^k$,
where $c_k$ are defined in terms of the matrix elements of the interactions 
and are independent of $\lambda$. In this case all terms in the series for 
$p(t)$ have the same order at $t_a$. However, this assumption is not true 
as $H_0$ and $V$ do not commute hence 
$\la 1 | e^{-iHt}|1\ra \neq \la 1 | e^{-i\lambda V t}|1\ra $.

We would like to mention a few interesting problems related to the Zeno 
effect. 1) A better characterization of the anti-Zeno region. This problem
is relevant to the experimental demonstration of the (anti-) Zeno behavior
of the survival probability. 2) How the non-ideal measurements influence
the Zeno effect? 3) Is the asymptotic quantum Zeno dynamics 
$\lim\limits_{N\to\infty} p_N(T)$ governed by a unitary group or a 
semigroup of isometries or contractions~\cite {Pasc3}? This question 
defines if the quantum Zeno dynamics introduces irreversibility in 
the evolution of a system.

\acknowledgments 
We would like to thank Profs. Ilya Prigogine, Tomio Petrosky and Saverio 
Pascazio for helpful discussions. We would also like to acknowledge the
remarks of Profs. Kofman and Kurizki. This work enjoyed the financial 
support of the European Commission Project No. IST-1999-11311 (SQID).

\appendix 
\section {Time evolution in the Heisenberg representation}

The second quantised form of the well-known Friedrichs model~\cite {Fried} 
is given by the Hamiltonian~(\ref{H}). For $\omega_1 > 0$ the oscillator 
excitations are unstable due to the resonance between the oscillator energy 
levels and the energy of a photon. Strong interaction however may lead to 
the emergence of a bound state. In weak coupling cases discussed here
bound states do not arise (see (\ref{ffb1})) below).

The solution of the eigenvalue problem
\equation\label{evp}
  [H, B^\dag_\omega]  =  \omega B^\dag_\omega
    \qquad \mbox{and} \qquad
  [H, B_\omega]  =  -\omega B_\omega,
\endequation
obtained with the usual procedure of the Bogolubov transformation
\cite {AGPP,KPPP} is
\begin{eqnarray} \label{bbp}
  (B^\dag_\omega)_{\stackrel{\rm in}{\rm out}}
    & = & b^\dag_\omega + \frac{\lambda f(\omega)}{\eta^\pm(\omega)}
    \int_{0}^{\infty}\!\!\!d\omega'\lambda f(\omega')
    \left(\frac{b^\dag_{\omega'}}{\omega'-\omega \mp i0} - a^\dag \right),\\
\label{bbm}
  (B_\omega)_{\stackrel{\rm in}{\rm out}}
    & = & b_\omega + \frac{\lambda f(\omega)}{\eta^\mp(\omega)}
          \int_{0}^{\infty}\!\!\!d\omega'\lambda  f(\omega')
          \left(\frac{b_{\omega'}}{\omega'-\omega \pm i0} - a \right).
\end{eqnarray}
In (\ref{bbp}), (\ref{bbm}) we used the notation
$ 1/\eta^\pm(\omega) \equiv 1/\eta(\omega \pm i0)$
where the function $\eta(z)$ of the complex argument $z$ is
\equation \label{etaz}
  \eta(z)
    = \omega_1-z-\int_0^\infty\!\!\! d\omega
            \frac{\lambda^2 f^2(\omega)}{\omega-z}.
\endequation
The following condition on the formfactor $f(\omega)$
\equation \label{ffb1}
  \omega_1 - \int_0^\infty\!\!\!d\omega\,
  \frac{\lambda^2 f^2(\omega)}{\omega} > 0
\endequation
guarantees that the function $1/\eta(z)$ is analytic 
everywhere on the first sheet of the Riemann manifold 
except for the cut $[0,\infty)$. Therefore the total Hamiltonian $H$
has no discrete spectrum and there are no bound states.

The incoming and outgoing operators 
$(B^\dag_\omega)_{\stackrel{\rm in}{\rm out}}$,
$(B_\omega)_{\stackrel{\rm in}{\rm out}}$ 
satisfy the following commutation relation
\equation \label{crbb}
  \left[(B_\omega)_{\stackrel{\rm in}{\rm out}},
        (B^+_{\omega'})_{\stackrel{\rm in}{\rm out}}\right]
    = \delta(\omega-\omega').
\endequation
The other commutators vanish. 
The bare vacuum state $|0\rangle$ satisfying
$$
  a_1|0\rangle = b_\omega|0\rangle = 0,
$$
is also the vacuum state for the new operators:
$$
  (B_\omega)_{\stackrel{\rm in}{\rm out}}|0\rangle = 0 .
$$
Therefore, the new operators diagonalise the total Hamiltonian (\ref{H}) as
\equation \label{Hdiagonal}
  H = \int^{\infty}_0\!\!\!d \omega\,\omega\,
      (B^\dag_\omega)_{\stackrel{\rm in}{\rm out}}
      (B_\omega)_{\stackrel{\rm in}{\rm out}} .
\endequation
Using the inverse relations
\begin{eqnarray}\label{bplee}
 b^\dag_\omega
    & = & (B^\dag_\omega)_{\rm in} -  \lambda f(\omega)
          \int_0^{\infty}\!\!\! d\omega'
          \frac{\lambda f(\omega')}{\eta^-(\omega')}
          \frac{(B^\dag_{\omega'})_{\rm in}}{\omega'-\omega-i0}, \\[2ex]
\label{bmlee}
  b_\omega
    & = & (B_\omega)_{\rm in} - \lambda f(\omega)
          \int_0^{\infty}\!\!\! d\omega'
          \frac{\lambda f(\omega')}{\eta^+(\omega')}
          \frac{(B_{\omega'})_{\rm in}}{\omega'- \omega+i0}, \\ [2ex]
\label{aplee}
  a^\dag
    & = & - \int_0^\infty\!\!\! d\omega
            \frac{\lambda f(\omega)}{\eta^-(\omega)}
            (B^\dag_\omega)_{\rm in}, \\[2ex]
\label{amlee}
  a
    & = & - \int_0^\infty\!\!\! d\omega
            \frac{\lambda f(\omega)}{\eta^+(\omega)}
            (B_\omega)_{\rm in},
\end{eqnarray}
we obtain the time evolution of the bare creation and annihilation operators
in the Heisenberg representation:
\begin{eqnarray}
b^\dag_\omega(t)
    & = & b^\dag_\omega e^{i\omega t}  +  \lambda f(\omega)
          \left\{\int_0^{\infty}\!\!\! d\omega'\,\lambda f(\omega')
                 \frac{g(\omega',t) - g(\omega,t)}{\omega'-\omega}
                 b^\dag_{\omega'} - g(\omega,t)a^\dag
          \right\}, \nonumber \\[2ex]
  b_\omega(t)
    & = & b_\omega e^{-i\omega t}  +  \lambda f(\omega)
          \left\{\int_0^{\infty}\!\!\! d\omega'\,\lambda f(\omega')
                 \frac{g^*(\omega',t) - g^*(\omega,t)}{\omega'-\omega}
                 b_{\omega'}  - g^*(\omega,t)a
          \right\}, \nonumber \\[2ex]
  a^\dag(t)
    & = & \int_0^\infty\!\!\! d\omega\, \lambda f(\omega)
          g(\omega',t)b^\dag_{\omega'} + A(t)a^\dag ,
          \nonumber \\[2ex]
\label{abt}
  a(t)
    & = & \int_0^\infty\!\!\! d\omega\, \lambda f(\omega)
          g^*(\omega',t)b_{\omega'} + A^*(t)a .
\end{eqnarray}
Except for the oscillating exponent,
all time dependence of the field operators is
described by the functions $g(\omega,t)$ and $A(t)$:
\equation \label{g}
  g(\omega,t)  = 
- \frac{1}{2\pi i}\int_{-\infty}^\infty \! d\omega'\frac{1}{\eta^-(\omega')}
        \frac{e^{i\omega' t} }{\omega' - \omega - i0},
\endequation
\equation \label{A}
  A(t)
    =\left(i\frac{\partial}{\partial t} + \omega\right)g(\omega,t)
    =\frac{1}{2\pi i}\int_{-\infty}^\infty \!\!\! d\omega'
     \frac{e^{i\omega' t}}{\eta^-(\omega')}.
\endequation

\section {Type 1 formfactor}

For the formfactor $\varphi_1(x)={\sqrt{x}\over 1+x}$ we have:
\beq \label {etaex}
  \eta_\Lambda(z)
    = \omega_{\Lambda}-z-\frac{\pi\lambda^2}{1-i\sqrt{z}},
\eeq
where the first sheet of the complex $z$ plane corresponds to
the upper half of the complex $\sqrt{z}$ plane.
The exact expression for the survival amplitude is 
known~\cite {photodetachement1,kofman-opt,LP}:
\beq
  A(t)
    = {i \gamma +\sqrt{\tilde\omega_\Lambda}
         \over 2\gamma\sqrt{\tilde\omega_\Lambda}}
     {\pi \lambda^2 \over z_3-z_2} e^{iz_2\Lambda t}
    + \pi e^{i\pi \over 4}\lambda^2
    \sum_{k=1}^3 \left( \prod_{\stackrel{m=1}{m\neq k}}^3
{1 \over z_k-z_m} \right) \sqrt{iz_k} e^{iz_k \Lambda t}
\left(-1+{\rm erf}(\sqrt{iz_k\Lambda t})\right).
\label {basexp}
\eeq
Here $z_k$ are the roots of $\eta_\Lambda(z)$ on the second sheet 
of $z$-plane, and $\tilde\omega_\Lambda$, $\gamma$ are expressed 
in terms of $z_k$. 
If conditions~(\ref{labelA}) are satisfied, we have the following
approximate expressions:
\beq
\gamma \approx {\pi \over 2} \lambda^2, \qquad
\tilde\omega_\Lambda \approx \omega_\Lambda.
\label {paramrelations}
\eeq

In order to analyze the survival probability for large times, we need
the asymptotics of the $A(t)$ as $t\rightarrow\infty$:
$$
A(t) = { 2 i t^{-3/2}\over z_1 z_2 z_3}
\left(1-{12\over t} \sum_{k=1}^3 {1\over i z_k} +O(1/t^2)  \right).
$$
In the last expression, we can use the first term only when
$t \gg 12 \Big|\sum_{k=1}^3 {1\over i z_k}\Big| \approx {24\over \omega_1}$.
We have in fact checked numerically, that this is valid even on
shorter times.

Using (\ref{paramrelations}), we can now calculate the survival probability:
\beq
p(t) \approx  e^{-4\gamma t \sqrt{\omega_1\Lambda}}+
{\pi \lambda^4 \Lambda \over 4 \omega_1^4 t^3}-
{\sqrt{\pi} \lambda^2 \Lambda^{1/2} \over \omega_1^2 t^{3/2}} 
e^{-2\gamma t \sqrt{\omega_1\Lambda}} \cos{(\omega_1t - \pi/4)}
\quad\mbox{when}\quad t \gg {24\over \omega_1}.
\label{survlargetime}
\eeq
One can see that the survival probability decays exponentially for 
intermediate times, while for large times there is a power law.
We can calculate the transition time $t_{ep}$ when the exponential decay 
is replaced by the power law. This happens when these two terms in the 
expression for $p(t)$ are equal.  This condition leads to a transcendental
equation which can be approximately solved
$$
t_{ep} \approx - {5 \log{
\left((2\pi^4)^{0.4} \lambda^4 {\Lambda \over \omega_1}\right)} 
\over 4\pi\lambda^2\sqrt{\Lambda \omega_1}}.
$$
We should notice that in the vicinity of $t_{ep}$ the survival
probability oscillates with the frequency $\omega_1$.

Let us now discuss the asymptotics of the (\ref{basexp}) for small
times $t \sim 0$. From the definition of the survival probability $p(t)$ 
we know that $|A(0)|=1$. As the evolution is
unitary, we know that a linear term in the expansion of $p(t)$ vanishes 
in the vicinity of $t=0$. Expanding~(\ref{basexp}) at small times,
we find for the survival probability
\beq
p(t) = 1 - \left({t \over t_a}\right)^{1.5}
+\left({t \over t_b}\right)^2 +O(t^{5/2}).
\eeq
where $t_a = ( 3/(4\sqrt{2\pi}))^{2/3}/( \lambda^{4/3} \Lambda)$, 
and $t_b = 1/(\sqrt{\pi} \lambda \Lambda)$.

\section {Type 2 formfactor}

For the formfactor $\varphi_2(x)={x \over (1+x^2)^2}$ the dimensionless 
function $\eta_\Lambda(z)$ is
\beq
  \eta_\Lambda(z)
    = \omega_{\Lambda} -z - \lambda^2\frac{\pi - 2z}{4(1+z^2)}
    + \lambda^2\frac{\pi z^2+2z(\log z - i\pi)}{2(1+z^2)^2}
\eeq
This function has no roots on the first Riemann sheet. The roots on the 
second sheet are defined by the equation
\beq
\eta_\Lambda(z) + {2\pi i z \lambda^2 \over (1+z^2)^2} = 0.
\label {rootsA3}
\eeq
Inserting (C1) into (\ref{Adl}), we can see that the integrand 
vanishes at infinity at the upper half of the complex $z$ plane and we 
can change the contour of the integration as it is shown in Fig.~4.
Hence only two roots of~(\ref{rootsA3}) contribute to $A(t)$:
\begin {eqnarray*}
z_1 & = & \omega_1+i\frac{\gamma_1}{2}
\approx  \omega_\Lambda+i\pi\lambda^2 \omega_\Lambda, \\
z_2 & \approx & {\sqrt{\pi}\over 2} \lambda+i.
\end {eqnarray*}
It is interesting to notice that the root $z_2$ does not approach the 
continuous spectrum when $\lambda \to 0$. Instead, $z_2$ ``annihilates''
with the root $z_3 \approx -{\sqrt{\pi}\over 2} \lambda+i$, which however
does not contribute to the survival amplitude.

Combining the pole contributions with the background integral, we have
for the survival amplitude
\beq \label{A2}
  A(t)
    = \sum_{k=1}^2 R(z_k) e^{iz_k\Lambda t}
   + \lambda^2 \int\limits_0^\infty dx
      \frac{x(1-x^2)^2e^{-x\Lambda t}}
           {(Q(x)+\frac{1}{2}\lambda^2\pi x)(Q(x)-\frac{3}{2}\lambda^2\pi x)}
    = \sum_{k=1}^2 R(z_k) e^{iz_k\Lambda t} + \lambda^2 I(t),
\eeq
where
$$
Q(x)=(\omega_{\Lambda}-ix)(1-x^2)^2 - {\lambda^2 \over 4}(\pi-2ix)(1-x^2)
- {\lambda^2 \over 2}(\pi x^2 -2ix\log{x}),
$$
and
$$
  R(z) = - \left[1 - \frac{\lambda^2}{2}
        \left(\frac{3-z^2+2\pi z}{(1  + z^2)^2} 
        + \frac{1-3z^2}{(1+z^2)^3}(\pi z + 2\log z +2i\pi)
        \right) \right]^{-1}.
$$
It is worth noticing that we have two exponential terms in 
representation~(\ref{A2}). The first corresponds to the usual exponential 
decay of the system. The second decays very fast, with the time constant
$1/2\Lambda$. However, this term is very important for description of the
survival amplitude at times $t \sim 1/\Lambda$. As shown in Section~4,
in this region the Taylor expansion at $t=0$ already cannot be used,
hence the representation~(\ref{A2}) is the only way to get results.
We would like to notice that for the interaction $\varphi(x)={x \over 
(1+x^2)^4}$ there are three roots contributing to the survival amplitude:
$z_1 \approx \omega_\Lambda+i\pi\lambda^2 \omega_\Lambda$, $z_2 \approx 
i(1-\sqrt{\lambda}\sqrt[4]{\pi \over 8} e^{\pi i/8})$, and 
$z_3 \approx i(1-\sqrt{\lambda}\sqrt[4]{\pi \over 8} e^{5\pi i/8})$. 
Hence, the expressions for the survival amplitude previously 
obtained~\cite {P1,P2} cannot be used for 
arbitrary time $t$ and should be corrected for $t \sim 1/\Lambda$ 
with adding two additional exponential terms.

Let us calculate first the long-time asymptotics.
For the integral term in the $A(t)$ we have
$$
I(t) = {1\over Q^2(0)\Lambda^2 t^2 }
\left( 1+ {4i\over t \Lambda Q(0)} +O(1/(\Lambda t)^2)\right).
$$
As in Appendix~B, we can use only one term of the asymptotics when
$ t \gg 4/\omega_1$. In this region, the survival probability can 
be written as
\beq
p(t)\approx e^{-\gamma_1 \Lambda t} + {\lambda^4\over Q^4(0)\Lambda^4 t^4 }-
{2 \lambda^2 e^{-\gamma_1 \Lambda t/2}\over Q^2(0)\Lambda^2 t^2 }
\cos{(\omega_1 t)}.
\eeq
Here again we can see two regions: intermediate with exponential behaviour 
and long tail with the power law decay. The transition time
$t_{ep}$ can also be calculated:
$$
t_{ep} = - {4\over \gamma_1} \log{ \lambda \gamma_1 \over Q(0)\Lambda}.
$$

In order to calculate the short-time asymptotics we expand $I(t)$ into 
the series at $t=0$:
\beq
I(t) \approx C_0 + C_1 t + C_2 t^2 + C_3 t^3 + 
\int\limits_0^\infty dx
      \frac{x (-x\Lambda)^4(1-x^2)^2e^{-x \Lambda t}}
      {(Q(x)-\frac{1}{2}\lambda^2\pi x)(Q(x)+\frac{3}{2}\lambda^2\pi x)},
\eeq
where $C_i$ are constants. The asymptotics of the integral term in the last 
expression can be easily found~\cite {IntegralTransformation}:
$$
I^{(4)}(t) \approx -\Lambda^4\int\limits_0^\infty dx 
{e^{-x \Lambda t} \over x+2i\omega_\Lambda} =
\Lambda^4 \log{(2i\omega_\Lambda t)} +O(1).
$$
Combining these results with Eq.~(\ref{p-expansion}), we get
\beq
p(t)=1-\left({t \over t_a}\right)^2-{\lambda^2 \over 12}\log{(2\omega_1 t)}
\Lambda^4 t^4+O(t^4),
\eeq
where $t_a={\sqrt{2} \over \lambda\Lambda}$.

\newpage
\begin {center}{\bf Figure captions} \end {center}

Fig. 1. The survival probability $p(t)$ for the photodetachement model
($\varphi_1(x)={\sqrt{x} \over 1+x}$, the dashed line), and for the
quantum dot model ($\varphi_2(x)={x \over (1+x^2)^2}$, the solid line).
The Zeno time $t_Z$ is indicated. Time is in units of the decay time $t_d$.

Fig. 2. The probability $p_N(T)$ (Eq.~(\ref{pnt})) as a function of 
the duration $\tau$ between measurements. From above, the curves correspond 
to the time of observation $T = 10^{-4}$, $10^{-3}$, $10^{-2}$, and 
$10^{-1}$, respectively. $T$ and $\tau$ are in units of the decay time 
$t_d$. The photodetachement model ($\varphi_1(x)={\sqrt{x} \over 1+x}$)
(Fig.~2a) and the quantum dot model ($\varphi_2(x)={x \over (1+x^2)^2}$)
(Fig.~2b) are presented.

Fig. 3. The value $N_\varepsilon(T)$ (Eq.~(\ref{Neps})) as a function 
of observation time $T$. From above, the curves correspond to 
the accuracy $\varepsilon = 10^{-2}$, $3\ 10^{-3}$, and $10^{-3}$,
respectively. The solid lines are for the photodetachement model
($\varphi_1(x)={\sqrt{x} \over 1+x}$), and the dashed lines are for the 
quantum dot model ($\varphi_2(x)={x \over (1+x^2)^2}$). $T$ is in units 
of the decay time $t_d$.

Fig. 4. The contour of integration.

\newpage

Table 1. The Zeno time $t_Z$, the time $t_a$, the decay time $t_d$,
and the time $t_{ep}$ of the transition from the exponential to power law 
decay for different model of interactions and for different physical 
systems. Numerical values are given in seconds and in units of $t_d$.

\begin{tabular}{cccc} \hline
formfactor $\varphi(x)$ & $\sqrt{x} \over 1 + x $ &
$x \over \left(1 + x^2 \right)^2 $ & 
$x \over \left(1 + x^2 \right)^4 $ \\ \hline
$t_Z$ & ${32\over 9\pi}{1\over\Lambda}$ & 
${\sqrt{6}\over 
\Lambda\sqrt{|\log{({2\sqrt{6}\omega_1 \over \Lambda})}|}}$ &
$ 2\sqrt{6} \over \Lambda$ \\
$t_a$  & $({3\over 4\sqrt{2\pi}})^{2/3}\over \lambda^{4/3}\Lambda$
& ${\sqrt{2} \over \lambda\Lambda}$ & $\sqrt{6} \over \lambda\Lambda $ \\
$t_d$ & $ {1\over \pi\lambda^2 \sqrt{\Lambda\tilde\omega_1}}$ &
${1\over 2\pi\lambda^2 \omega_1}$ & ${1\over 2\pi\lambda^2 \omega_1}$ \\
$t_{ep}$ & $- {5 \log({\lambda^4 {\Lambda \over 
\tilde\omega_1})} \over 4\pi\lambda^2\sqrt{\Lambda \tilde\omega_1}}$ 
& $-{2\log{(2\pi\lambda^3)} \over \pi\lambda^2 \omega_1}$
& $-{2\log{(2\pi\lambda^3)} \over \pi\lambda^2 \omega_1}$ \\ \hline
system & photodetachement & quantum dot & hydrogen atom \\ \hline
$\Lambda$, s$^{-1}$ & $1.0 \ 10^{10}$ & $1.67 \ 10^{16}$ & 
 $8.498 \ 10^{18}$ \\
$\omega_1$, s$^{-1}$ & $2.0 \ 10^{4}$ & $7.25 \ 10^{12}$ & 
 $1.55 \ 10^{16}$ \\
$\lambda^2$ & $3.18 \ 10^{-7}$ & $3.58 \ 10^{-6}$ & 
$6.43 \ 10^{-9}$ \\ \hline
$t_Z$, s ($t_d$) & $1.1 \ 10^{-10}$ ($1.1 \ 10^{-9}$) & 
$5.9 \ 10^{-17}$ ($9.7 \ 10^{-9}$) & $5.76 \ 10^{-19}$ ($3.6 \ 10^{-10}$) \\
$t_a$, s ($t_d$) & $9.6 \ 10^{-7}$ ($9.6 \ 10^{-6}$) & 
$4.5 \ 10^{-14}$ ($7.4 \ 10^{-6}$) & $3.59 \ 10^{-15}$ ($2.2 \ 10^{-6}$) \\
$t_d$, s ($t_d$) & 0.1 (1) & $6.1 \ 10^{-9}$ (1) & $1.60 \ 10^{-9}$ (1) \\
$t_{ep}$, s ($t_d$) & 
1.7 (17)& $4.2 \ 10^{-7}$ (69) & $1.69 \ 10^{-7}$ (110)\\
\hline
\end{tabular}

\end {document}
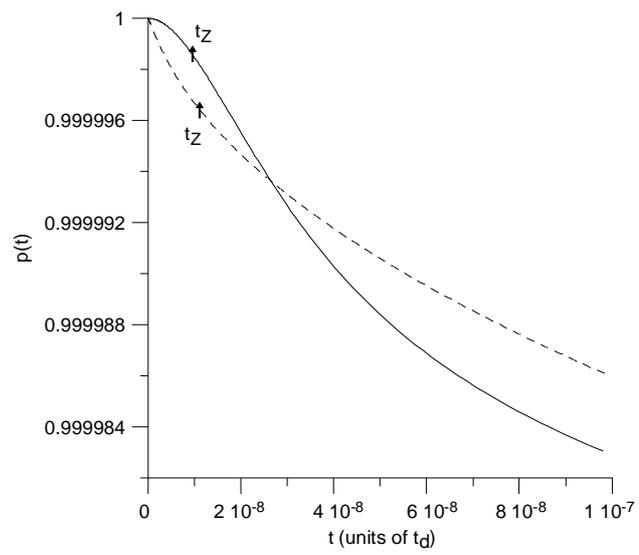
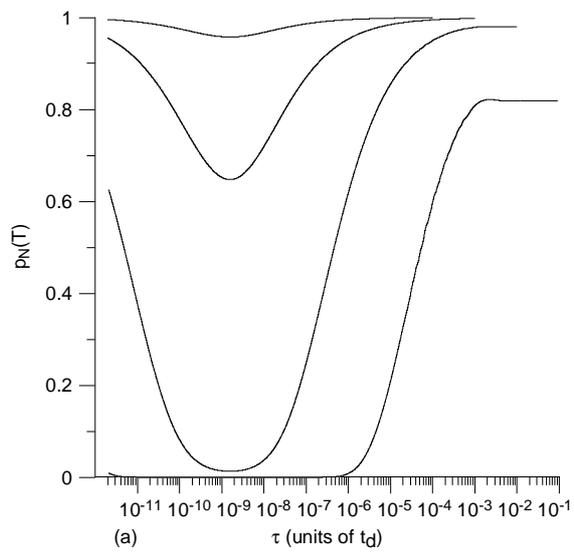
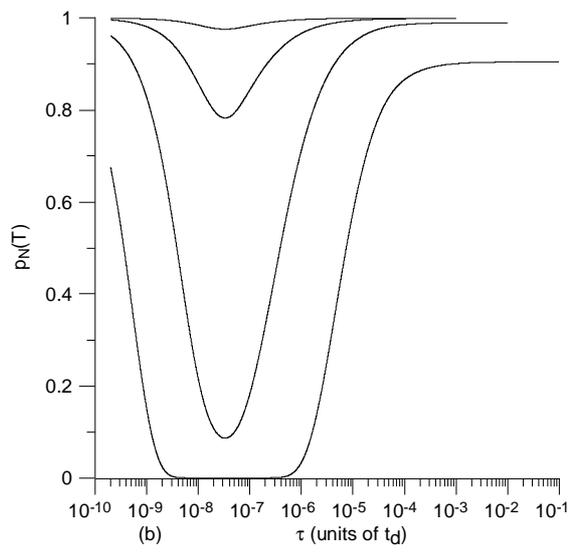
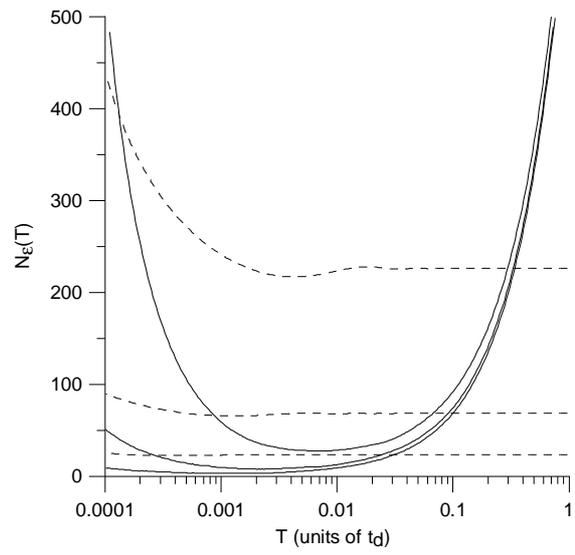
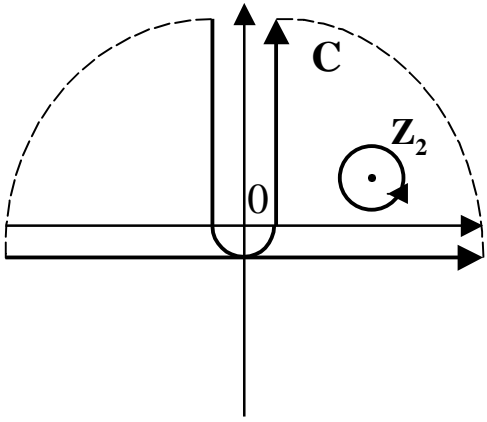
